\documentclass[10pt,twocolumn]{IEEEtran}

\usepackage{setspace}
\usepackage{clrscode}
\usepackage{amsmath}
\usepackage{amssymb}
\usepackage[dvips]{graphicx}
\usepackage{epsfig}
\usepackage{hyperref}
\usepackage{bm}
\usepackage{subfigure}
\usepackage{verbatim}
\usepackage{titletoc}
\usepackage{extarrows}
\usepackage{fancyhdr}
\usepackage{booktabs}
\usepackage{mathrsfs}
\usepackage{color}
\usepackage{amssymb,mathrsfs,amsmath}
\usepackage{bbding}
\usepackage{microtype}

\usepackage{lipsum}
\usepackage{cuted}\stripsep-3pt
\usepackage{stfloats}

\usepackage{algorithm}
\usepackage{algorithmic}

\usepackage{balance}

\usepackage{cite}

\newcommand{\teq}{\triangleq}
\newcommand{\mbf}{\mathbf}
\newcommand{\mbb}{\mathbb}
\newcommand{\mcal}{\mathcal}
\newcommand{\tbf}{\textbf}

\newcommand{\tr}{\text{Tr}}

\newcommand{\bs}{\boldsymbol}
\newcommand{\IRS}{\bm{\Phi}}
\newcommand{\bg}{\bs{g}}
\newcommand{\bh}{\bs{h}}

\newcommand{\bn}{\bs{n}}
\newcommand{\A}{\text{A}}
\newcommand{\F}{\text{F}}
\newcommand{\dia}{\text{diag}}

\newcommand{\noi}{\sigma^2}

\newcommand{\bQ}{\mbf{Q}}
\newcommand{\bB}{\mbf{B}}

\newcommand{\bV}{\mbf{V}}
\newcommand{\bW}{\mbf{W}}

\newcommand{\E}{\text{E}}
\newcommand{\bb}{\bs{b}}
\newcommand{\bv}{{\bf{v}}}

\newcommand{\dv}{\hat{{\bv}}}

\newcommand{\pset}{\{p_k\}}

\newcommand{\vset}{\{\bv_k\}}
\newcommand{\fset}{\{f_k\}}

\newcommand{\chiset}{\{\chi_k\}}
\newcommand{\ioset}{\{\iota_k\}}
\newcommand{\rank}{\text{rank}}

\newcommand{\tQ}{\Tilde{\bQ}}
\newcommand{\pro}{{\mathcal{P}}}
\newcommand{\nareq}{\hspace{-0.1cm}=\hspace{-0.1cm}}

\newenvironment{sequation}{\begin{equation}\small}{\end{equation}\noindent}

\newenvironment{narsequation}{\small\begin{equation}\setlength\abovedisplayskip{0.15cm}\setlength\belowdisplayskip{0.15cm}}{\end{equation}\noindent}

\newtheorem{Rem}{\tbf{Remark}}

\begin{document}

\title{ { {Throughput Maximization for Active Intelligent Reflecting Surface Aided Wireless Powered Communications}}}
    \author{
		\IEEEauthorblockN{Piao Zeng, Deli Qiao, Qingqing Wu, \emph{Senior Member, IEEE}, and Yuan Wu } 
		\thanks{P. Zeng and D. Qiao are with the School of Communication and Electronic Engineering, East China Normal University (e-mail: 52181214005@stu.ecnu.edu.cn; dlqiao@ce.ecnu.edu.cn).  
		Q. Wu and Y. Wu are with the State Key Lab of Internet of Things for Smart City, of University of Macau (email: qingqingwu@um.edu.mo; yuanwu@um.edu.mo). }
	}

\maketitle

	\begin{abstract}
		This paper considers an active intelligent reflecting surface (IRS)-aided wireless powered communication network (WPCN), where devices first harvest energy and then transmit information to a hybrid access point (HAP). Different from the existing works on passive IRS-aided WPCNs, this is the first work that introduces the active IRS in WPCNs. To guarantee fairness, the problem is formulated as an amplifying power-limited weighted sum throughput (WST) maximization problem, which is solved by successive convex approximation technique and fractional programming alternatively. To balance the performance and complexity tradeoff, three beamforming setups are considered at the active IRS, namely user-adaptive IRS beamforming, uplink-adaptive IRS beamforming, and static IRS beamforming. 
		Numerical results demonstrate the significant superiority of employing active IRS in WPCNs and the benefits of dynamic IRS beamforming. Specifically, it is found that compared to the passive IRS, the active IRS not only improves the WST greatly, but also is more energy-efficient and can significantly extend the transmission coverage. Moreover, different from the symmetric deployment strategy of passive IRS, it is more preferable to deploy the active IRS near the devices.
	\end{abstract}
	\begin{IEEEkeywords}
		 Active intelligent reflecting surface, dynamic beamforming, wireless powered communication.
	\end{IEEEkeywords}
	
\section{Introduction}
Intelligent reflecting surface (IRS) has recently been intensively investigated to enhance the performance of wireless powered communications, such as the joint optimization in the IRS-assisted simultaneous wireless information and power transfer (SWIPT) system \cite{pan2020intelligent} and IRS-aided wireless powered communication networks (WPCNs) \cite{zheng2020intelligent,wu2021irs,zou2020wireless}. Besides, there are also works considering a novel self-sustainable IRS architecture, where the IRS can work in both energy harvesting and signal reflecting phases \cite{zou2020wireless,ntontin2021toward}. 
However, despite it is shown in \cite{wu2019intelligent} that IRS can bring an asymptotic squared power gain, the actual capacity gains for an IRS are negligible when the direct link is not weak due to the ``double fading'' path loss effect in IRS-aided cascaded channel \cite{najafi2020physics,tang2020wireless,long2021active}. 
To overcome the double-fading issue of the  conventional passive IRS, a new type of IRS, i.e., active IRS, has recently been proposed, which is able to amplify the reflected signal's amplitude \cite{long2021active,zhang2021active,you2021wireless}.

Unlike the multi-antenna amplify-and-forward (AF) relay that requires power-consuming radio-frequency (RF) chains and two time slots to complete the AF processing, the proposed active IRS basically inherits the hardware structure of the conventional passive IRS except replacing the passive reflecting elements (REs) with the active load impedances, which can directly reflect the incident
signal with the desired adjustment in the electromagnetic level and amplify the signals in the air without the reception \cite{long2021active,zhang2021active}.

Although the active IRS-aided system presents a desirable performance in regard to signal transmission \cite{long2021active,zhang2021active}, whether this advantage still holds or not when employing the active IRS in WPCNs compared with its passive counterpart is unknown since the active IRS requires additional amplifying power and the energy consumption is a crucial issue in WPCNs.
On the other hand, as revealed in \cite{wu2021irs}, for passive IRS-aided WPCNs, the UL-adaptive scheme is equivalent to the static IRS beamforming scheme. However, for active IRS, the problem formulation is rather different due to the newly imposed amplifying constraint. Therefore, whether the aforementioned conclusion still holds remains an open question in active IRS aided WPCNs.

Motivated by the above, in this paper we investigate an active IRS aided WPCN where an active IRS is deployed to assist the downlink (DL) wireless energy transmission (WET) and uplink (UL) wireless information transmission (WIT) between a hybrid access point (HAP) and multiple devices, as shown in Fig. \ref{fig:Channel}. To the best of our knowledge, this is the first work that introduces the active IRS in WPCNs. To guarantee fairness, we aim to maximize the weighted sum throughput (WST) of all devices via jointly optimizing the resource allocation and the reflection coefficient matrix, subject to the amplification power and amplifying amplitude constraints at the active IRS. 
In particular, to trade-off between the practical performance and implementation complexity, we investigate three beamforming setups at the active IRS, namely user-adaptive IRS beamforming, UL-adaptive IRS beamforming, and static IRS beamforming, as elaborated later.

Note that different from the work that investigated SWIPT system \cite{pan2020intelligent}, we consider an active IRS-aided WPCN, which is also different from the existing works on conventional passive IRS-aided WPCNs \cite{zheng2020intelligent,zou2020wireless,wu2021irs}. In particular, by employing the active IRS in the transmissions, the amplitude of the incident signal can be amplified, which brings new degrees of freedom in the design and adds the variables to be optimized. Besides, the additional constraints of amplification power and the maximum amplifying amplitude limitation at the active IRS further increase the difficulty and complexity for IRS beamforming optimization. 
Moreover, due to the use of active REs at the IRS, the thermal noises at the IRS are also amplified and thereby cannot be ignored as that in passive IRS-aided WPCNs. Specifically, in DL WET, the wireless-powered devices can harvest the additional energy from the amplified noise power; whereas in UL WIT, to maximize the communication throughput, the active REs’ reflecting coefficients have to balance the two conflicting goals, i.e., the received signal power maximization and the noise power minimization. 
Therefore, the joint design of resource allocation and IRS beamforming in the considered active IRS aided WPCNs is much more challenging than its counterpart in conventional passive IRS-aided WPCNs, thus calling for non-trivial efforts.

To solve the formulated challenging WST maximization problem, we decompose the coupled joint optimization into two subproblems and propose an alternating optimization (AO)-based algorithm, where successive convex approximation (SCA) technique and fractional programming (FP) are utilized to solve the two subproblems respectively\footnote{ { {Note that the proposed algorithm can be extended to solve the optimization with the hybrid IRS after slight modifications \cite{nguyen2021spectral}.} }}. 
Numerical results reveal that employing the active IRS in WPCNs performs much better than its passive IRS counterpart and validate the advantage of dynamic IRS beamforming, which is different from the conclusion in \cite{wu2021irs}. 
Specifically, it is found that: 1) The active IRS can improve the WST significantly by alleviating the double-fading effects suffered by the passive IRS; 2) Even less transmit energy and amplifying energy is required to achieve the higher throughput and/or the broader coverage with the assist of the active IRS; 
{\textls[0]{3) Different from the symmetric deployment strategy for passive IRS, it achieves much higher WST by deploying the active IRS closer to the devices rather than closer to the HAP.}}

 { {
	\emph{Notations:} Throughout the paper, superscripts $(\cdot)^{T}$, $(\cdot)^{H}$ and \text{diag}$(\cdot)$ represent the transpose,  Hermitian transpose and diagonalization operator, respectively. $\mbb{C}^{a \times b}$ denotes the space of $a \times b$ complex matrices. \tr$(\mbf{X})$ denotes the trace of the matrix $\mbf{X}$. $\Re\{\cdot\}$ represents the real part of a complex value. }}

\section{System Model}

\begin{figure}
		\centering
		\vspace{-0.4cm}
		\includegraphics[width=0.7\columnwidth]{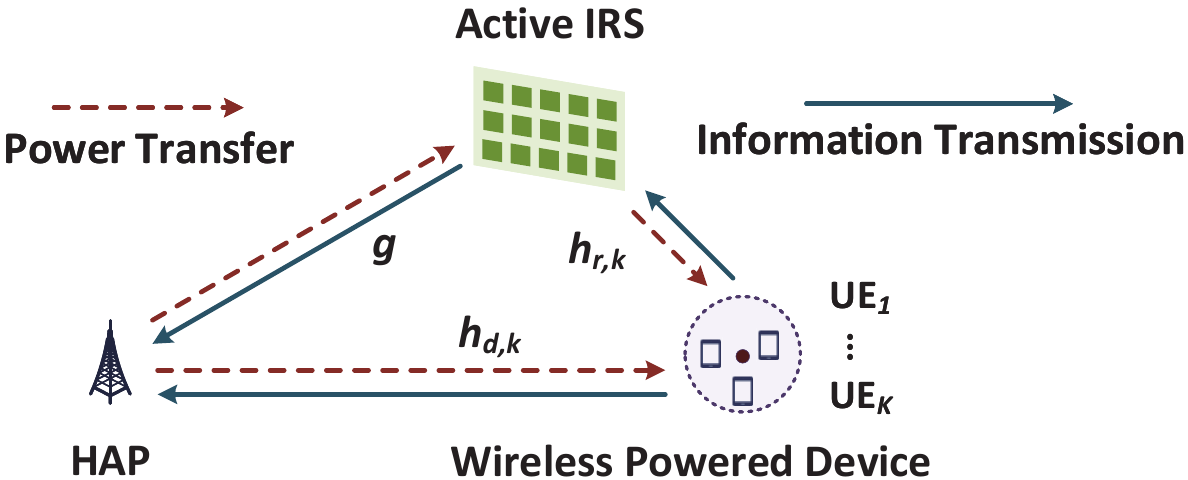}
		\caption{An active IRS-aided WPCN.}
		\label{fig:Channel}
\end{figure}

As shown in Fig. \ref{fig:Channel}, we consider an active IRS-aided WPCN, which consists of one HAP, one IRS and $K$ wireless-powered devices. We assume that the HAP and all the devices are equipped with a single antenna, while the IRS has $N$ active REs.
The equivalent baseband channels from the HAP to the IRS, from the IRS to device $k$, and from the HAP to device $k$ are denoted by $\bg \in \mbb{C}^{N \times 1}$, $\bh_{r,k} \in \mbb{C}^{N \times 1}$ and $h_{d,k} \in \mbb{C}$, $\forall k \in \mcal{K}\teq\{1,\cdots,K\}$, respectively. 
We assume that all the channels in the system follow the quasi-static flat-fading model,  { {
and the instantaneous channel state information (CSI) for all links is assumed to be available at the devices and the IRS, which can be obtained with the compressive sensing technique \cite{chen2019channel} or by the active sensors deployed at the IRS \cite{taha2021enabling}. }}

During DL WET, the HAP broadcasts the energy signal with a constant transmit power $P_\A$  for time $\tau_0$. Thus, the received signal at device $k$ is written as\footnote{ { {For the comparison with the AF relay-aided system, the readers may refer to \cite{wu2019intelligent} and \cite{ntontin2019multi} for more details.} }}
\begin{sequation}
		\begin{aligned}
			y_k = &\hspace{0.1cm}\sqrt{P_\A} ( h_{d,k}^{H} + \bh_{r,k}^{H} \IRS_0 \bg ) x_0 + \bh_{r,k}^{H} \IRS_0 \bn_1 + z_{1,k} \\
			=  &\hspace{0.1cm}\sqrt{P_\A} ( h_{d,k}^{H} + \bb_{k}^{H} \bv_0  ) x_0 + \bh_{r,k}^{H} \IRS_0 \bn_1 + z_{1,k},
		\end{aligned}
\end{sequation}\noindent
where {\small $\bb_{k}^H \teq \bh_{r,k}^H \dia(\bg)$. $\IRS_0 = \text{diag}(\bv_0)$} represents the diagonal reflection coefficient matrix for DL WET, with each diagonal entry being the corresponding entry in $\bv_0 \teq [\phi_{0,1},\cdots,\phi_{0,N}]^T$. 
Different from the existing passive IRS architecture, we consider that the IRS is supported by an external power source, and thus the REs in the IRS can exploit the active loads to amplify the incident signal so as to alleviate the double-fading attenuation \cite{long2021active,zhang2021active,you2021wireless}. In particular, the reflecting coefficient of the $n$-th RE at the IRS is denoted by  $\phi_{0,n}=a_{0,n}e^{j\theta_{0,n}},{\small \forall n  \in \mathcal{N}\teq\{1,\cdots,N\}} $, where $a_{0,n}\in [0,a_\text{max}]$ and ${\small \theta_{0,n} } \in [0,2\pi)$ represent the amplitude and the phase-shift (PS) of $\phi_{0,n}$, respectively \cite{long2021active}.
The amount of harvested energy at device $k$ during DL WET is thereby expressed as
\begin{sequation}
	\begin{aligned}
		\E_k ( \bv_0 )
		=& \hspace{0.1cm} \tau_0 \eta \big( P_\A | h_{d,k}^{H} + \bb_{k}^{H} \bv_0  |^2+ \noi_{n_1} \bv_0^H \bQ_{2,k} \bv_0  \big),
	\end{aligned}
\end{sequation}\noindent
where {\small $\bQ_{2,k} \teq \dia(|[\bh_{r,k}]_1|^2, \cdots,|[\bh_{r,k}]_N|^2)$} and {\small $\eta \in (0,1]$} is the energy conversion efficiency of the devices. 

{\textls[0]{ For UL WIT, each device transmits its own information signal to the HAP for a duration of $\tau_k$ with transmit power $p_k$. In the following, we investigate three cases of IRS beamforming setups, namely the user-adaptive IRS beamforming, UL-adaptive IRS beamforming and static IRS beamforming, depending on how the active IRS reconfigures its REs over time during UL WIT.} }

\section{Problem Formulation}
To guarantee fairness, we aim at maximizing the WST of the considered WPCN by jointly optimizing the IRS coefficient matrix, the time allocation and the devices' transmit power.

1) {\textls[0]{For the user-adaptive IRS beamforming case, the active IRS is allowed to reconfigure its reflecting vectors $K$ times in UL WIT and each vector is dedicated to one device. }}The achievable throughput of the $k$-th device in bits/Hz is computed as 
\begin{sequation}
	\begin{aligned}
		r_k(\tau_k, p_k,\bv_k) &= \tau_k \log_2 \bigg( 1+ \frac{ p_k | h_{d,k}^{H} + \bb_{k}^{H} \bv_k |^2 }{\noi_{n_{2}}  \bv_k^H \bQ_{1} \bv_k  + \noi_{z_{2,k}}} \bigg) ,
	\end{aligned}
\end{sequation}

{\begin{spacing}{1.05} \noindent
where {\small $\bQ_{1} \teq \dia(|[\bg]_1|^2, \cdots,|[\bg]_N|^2)$. $p_k$} and {\small $\bv_k \teq [\phi_{k,1},\cdots,$ $\phi_{k,N}]^H$}  are the transmit power and the IRS beamforming vector of the $k$-th device during $\tau_k$ in UL WIT, respectively. The problem can be accordingly formulated as \end{spacing}}
\begin{small}
		\begin{align}
		&({\pro_{UE}}):{ \max_{\scriptstyle \tau_0,\{\tau_k\},\atop \pset,  \scriptstyle \bv_0,\vset} }  \hspace{0.3cm}
			\sum\nolimits_{k=1}^K \omega_k r_k(\tau_k, p_k,\bv_k) \label{pro:A}
			\end{align} 
\end{small}\noindent
\vspace{-0.5cm}
\begin{small}
		\begin{align}
			&  \hspace{0.2cm} {\text{s.t.}} \hspace{0.2cm} { p_k \tau_k \leq \tau_0 \eta \big( P_\A | h_{d,k}^{H} + \bb_{k}^{H} \bv_0  |^2+ \noi_{n_1} \bv_0^H \bQ_{2,k} \bv_0  \big), \forall k, } \tag{\ref{pro:A}{a}} \label{pro:Aa} \\
			& \hspace{0.75cm} P_\A \bv_0^H \bQ_{1} \bv_0 + \noi_{n_1} \bv_0^H \bv_0 \leq P_\F  , \tag{\ref{pro:A}{b}} \label{pro:Ab} \\
			& \hspace{0.8cm} p_k \bv_k^H \bQ_{2,k} \bv_k + \noi_{n_2} \bv_k^H \bv_k \leq P_\F ,\forall k , \tag{\ref{pro:A}{c}} \label{pro:Ac} \\
			& \hspace{0.8cm} \tau_{0}+\sum\nolimits_{k=1}^K \tau_{k} \leq T_{\max }  , \tag{\ref{pro:A}{d}} \label{pro:Ad} \\
			& \hspace{0.8cm} \tau_{0} \geq 0, \tau_{k} \geq 0, p_k \geq 0, \forall k , \tag{\ref{pro:A}{e}} \label{pro:Ae} \\
			& \hspace{0.85cm} a_{t,n} \leq a_{\max},\hspace{0.2cm} t=0,1,\cdots,K, \hspace{0.2cm} \forall n\in \mcal{N} . \tag{\ref{pro:A}{f}} \label{pro:Af} 
		\end{align} 
\end{small}\noindent
In $({\pro_{UE}})$, (\ref{pro:Aa}) represents the energy causality constraints. (\ref{pro:Ab}) and (\ref{pro:Ac}) are the IRS amplifying power constraints for DL WET and UL WIT, respectively. (\ref{pro:Ad}) is the total time constraint. (\ref{pro:Ae}) are the non-negativity constraints.  (\ref{pro:Af}) are the amplitude constraints of the active REs. It can be seen that, by employing the active IRS, the amplitude and PS of the reflecting coefficient vectors $\bv_0$ and $\vset$ need to be optimized together, which are closely coupled with $\tau_0,\{\tau_k\}$ and $\pset$ in the objective as well as in the constraint (\ref{pro:Aa}) and (\ref{pro:Ac}). Therefore, 
problem (\ref{pro:A}) is a challenging non-convex optimization problem and difficult to be solved optimally in general.


{\begin{spacing}{1.05} 2) For UL-adaptive IRS beamforming case, all the devices share the common IRS coefficient vector in UL WIT, i.e., $\bv_1 = \cdots =\bv_K$. The corresponding problem can be formulated similarly to $(\pro_{UE})$, which is given by \end{spacing}}
\vspace{-0.2cm}
\begin{small}
		\begin{align}
		&({\pro_{UL}}): { \max_{\scriptstyle \tau_0,\{\tau_k\}, \atop \scriptstyle \pset,\bv_0,\bv_1} }  \hspace{0.2cm}
			\sum\nolimits_{k=1}^K \omega_k r_k(\tau_k, p_k,\bv_1) \label{pro:A2}\\
			&  \hspace{1.6cm} {\text{s.t.}} \hspace{0.8cm}  p_k \bv_1^H \bQ_{2,k} \bv_1 + \noi_{n_2} \bv_1^H \bv_1 \leq P_\F ,\forall k , \tag{\ref{pro:A2}{a}} \label{pro:A2a} \\
			& \hspace{2.8cm}  a_{t,n} \leq a_{\max},\hspace{0.2cm} t=0,1, \hspace{0.2cm} \forall n\in \mcal{N} ,\tag{\ref{pro:A2}{b}} \label{pro:A2b} \\
			& \hspace{2.7cm} (\ref{pro:Aa}),(\ref{pro:Ab}),(\ref{pro:Ad}),(\ref{pro:Ae}).\tag{\ref{pro:A2}{c}} \label{pro:A2c} 
		\end{align}
\end{small}\noindent

{\begin{spacing}{1.05}
3) For static IRS beamforming case, all the devices need to share the same IRS coefficient vector as that in DL WET, i.e., $\bv_0 \hspace{-0.1cm}= \hspace{-0.1cm} \bv_1 \hspace{-0.1cm}= \hspace{-0.1cm} \cdots \hspace{-0.1cm}= \hspace{-0.1cm} \bv_K$. Accordingly, the problem is formulated as \end{spacing}}
\vspace{-0.2cm}
\begin{small}
		\begin{align}
	    \hspace{-2.5cm}	({\pro_{ST}}): { \max_{\scriptstyle \tau_0,\{\tau_k\},  \pset,\bv_0} }  \hspace{0.2cm}
			\sum\nolimits_{k=1}^K \omega_k r_k(\tau_k, p_k,\bv_0)  \label{pro:A_ST}
		\end{align}
\end{small}\noindent
\vspace{-0.4cm}
\begin{small}
		\begin{align}
			  \hspace{1cm} {\text{s.t.}} \hspace{1.cm} & p_k \bv_0^H \bQ_{2,k} \bv_0 + \noi_{n_2} \bv_0^H \bv_0 \leq P_\F ,\forall k , \tag{\ref{pro:A_ST}{a}} \label{pro:A_STa} \\
			 \hspace{3cm} & a_{0,n} \leq a_{\max},\hspace{0.2cm} \hspace{0.2cm} \forall n\in \mcal{N} ,\tag{\ref{pro:A_ST}{b}} \label{pro:A_STb} \\
			 \hspace{2.4cm} &(\ref{pro:Aa}),(\ref{pro:Ab}),(\ref{pro:Ad}),(\ref{pro:Ae}).\tag{\ref{pro:A_ST}{c}} \label{pro:A_STc}
		\end{align}
\end{small}\noindent

\vspace{-0cm}
\begin{Rem}
Denote the optimal IRS beamforming vectors of $({\pro_{UL}})$ as $\bv_0^*$ and $\bv_1^*$. For passive IRS-aided WPCNs, it was shown in \cite{wu2021irs} that $\bv_0^* \! = \!\bv_1^*$. However, due to the new design flexibility and amplifying power/amplitude constraints imposed by the active IRS, this conclusion
does not hold anymore. Specifically, let us take the single device's case for illustration.  
{\textls[0]{ For the UL-adaptive IRS beamforming, to maximize the harvesting energy in DL WET, by utilizing the Cauchy-Schwarz inequality, we can derive the optimal $a_{0,n}$  as $a_{0,n}^*=\min\big\{ \frac{c_0}{|[g]_n| |[h_r]_n|}, a_{\max}\big\}$,
where $ c_0^2 \hspace{-0.3cm}=\hspace{-0.3cm} {{P_\F}/{\sum_{n=1}^N\big(  \frac{P_\A}{ |[h_r]_n|^2}  \hspace{-0.1cm} + \hspace{-0.1cm}  \frac{\noi_{n_1} }{|[g]_n|^2 |[h_r]_n|^2}\big)} }$.\vspace{0.1cm}
Similarly, to maximize the achievable throughput, the optimal $a_{1,n}$ is given by $a_{1,n}^* \hspace{-0.4cm}=\hspace{-0.3cm}  \min\big\{ \frac{c_1}{|[g]_n| |[h_r]_n|}, a_{\max}\big\}$, 
where $c_1^2\hspace{-0.4cm}= \hspace{-0.4cm}{P_\F}/ $ $  {\sum_{n=1}^N\big(  \frac{p}{ |[g]_n|^2} \hspace{-0.1cm} + \hspace{-0.1cm} \frac{\noi_{n_2} }{|[g]_n|^2 |[h_r]_n|^2}\big) }$. \vspace{0.1cm}
Obviously $a_{0,n}^* \hspace{-0.1cm} \ne \hspace{-0.1cm} a_{1,n}^*, \forall n \in \mcal{N}$. Therefore, we have $\bv_0^* \hspace{-0.1cm} \ne \hspace{-0.1cm} \bv_1^*$, which implies the necessity of UL-adaptive IRS beamforming in active IRS-aided WPCNs. }}
Simulation results also validate the additional gain of employing UL-adaptive IRS beamforming.
\end{Rem}

\vspace{-0.1cm}
\section{Proposed Solution for $(\pro_{UE})$}
\vspace{-0.0cm}
For the problem of employing user-adaptive IRS beamforming, i.e., $(\pro_{UE})$, to simplify the highly coupled non-convex optimization problem, the original problem  is first decomposed into two subproblems and then solved separately in an alternative manner.  

\vspace{-0.35cm}
\subsection{Optimizing $\tau_0,\{\tau_k\},\pset$ and $ \bv_0$ with the given $\{\bv_k\}$}
\vspace{-0.cm}
For a given set of $\{\bv_k\}$, the signal-to-interference-plus-noise ratio (SINR) of each device is a constant, which is denoted as 
\begin{spacing}{1.1} \noindent {\small{$
\gamma_k(\bv_k) \teq   | h_{d,k}^{H}+ \bb_{k}^{H} \bv_k |^2/(\noi_{z_{2,k}}   +\noi_{n_{2}}\bv_k^H \bQ_{1} \bv_k   )$}}. 
\textls[0]{ {{To proceed, we introduce a new set of variables {\small{$\fset$}}, where {\small{$f_k = p_k \tau_k, \forall k,$}} and {\small{$\bW_0 \hspace{-0.1cm} 
=\hspace{-0.1cm} \tau_0 \bV_0$}}, where {\small{$\bV_0 \hspace{-0.1cm} 
=\hspace{-0.1cm} \tilde{\bv}_0 \tilde{\bv}_0^H$}} and {\small{$\tilde{\bv}_0 \hspace{-0.1cm} 
=\hspace{-0.1cm} [\bv_0^H, 1]^H$}}, which needs to satisfy {\small{$\bW_0 \succeq {\bm{0}}$}} and rank{\small{$(\bW_0) \hspace{-0.1cm} 
=\hspace{-0.1cm}  1$}}. Note that {\small{$\tr( \bW_0)=\tau_0 (\bv_0^H \bv_0 -1)$}}. }}}Then the subproblem can be reformulated as 
\end{spacing}
\vspace{-0.3cm}
\begin{small}
		\begin{align}
			&\hspace{-1.8cm} { \max_{\scriptstyle \tau_0,\{\tau_k\},  \fset,  \bW_0} } \hspace{0.4cm}
			\sum\nolimits_{k=1}^K \omega_k \tau_k \log_2 \big[ 1+ {\gamma_k(\bv_k) f_k }/{\tau_k} \big] \label{pro:B1}
	\end{align}
\end{small}
\vspace{-0.3cm}
\begin{small}
\begin{align}
			& \hspace{0cm} {\text{s.t.}} \hspace{0.45cm}  { f_k+\eta\tau_0\noi_{n_1}\leq  \eta   \tr[(P_\A\bB_k+ \noi_{n_1}\tQ_{2,k})\bW_0], \forall k, } \tag{\ref{pro:B1}{a}} \label{pro:B1a}\\
			& \hspace{0.8cm} P_\A \tr(\tQ_{1} \bW_0) + \noi_{n_1} \tr( \bW_0) \leq \tau_0( P_\F+P_\A+\noi_{n_1}) , \tag{\ref{pro:B1}{b}} \label{pro:B1b} \\
			& \hspace{0.8cm} f_k \bv_k^H \bQ_{2,k} \bv_k +  \noi_{n_2} \tau_1 \bv_k^H \bv_k \leq \tau_1 P_\F ,\forall k , \tag{\ref{pro:B1}{c}} \label{pro:B1c} \\
			& \hspace{0.8cm} (\ref{pro:Ad}), \tau_{0} \geq 0, \tau_{k} \geq 0, f_k \geq 0, \forall k  , \tag{\ref{pro:B1}{d}} \label{pro:B1d} \\
			& \hspace{0.8cm} [\bW_0]_{n,n} \leq a_{\max}^2\tau_0,\hspace{0.2cm} \hspace{0.2cm} n=1,\cdots N , \tag{\ref{pro:B1}{e}} \label{pro:B1e} \\
			& \hspace{0.8cm} [\bW_0]_{N+1,N+1} = \tau_0 ,\hspace{0.2cm} \bW_0 \succeq {\bm{0}}, \hspace{0.2cm} \rank(\bW_0)=1, \tag{\ref{pro:B1}{f}} \label{pro:B1f} 
		\end{align}
\end{small}\noindent
where {\small $\tQ_{1} \teq \dia(|[\bg]_1|^2, \cdots,|[\bg]_N|^2, 1)$} and {\small $\tQ_{2,k} \teq \dia( $ $|[\bh_{r,k}]_1|^2, \cdots,|[\bh_{r,k}]_N|^2, 1)$}. It is not difficult to verify that by relaxing the rank one constraint in (\ref{pro:B1f}), problem (\ref{pro:B1}) becomes a convex semidefinite program (SDP) and can be solved with the interior-point algorithm. 
Note that to maintain a better convergence, Gaussian randomization is applied to reconstruct the rank-one solution with the obtained $\bW_0$ after the whole AO algorithm converges instead of executing in each iteration. 

\vspace{-0.2cm}
\subsection{Optimizing $\vset$ with the given $\tau_0,\{\tau_k\},\pset$ and $ \bv_0$}	
For this subproblem, the optimal $\vset$ can be independently obtained by solving $K$ subproblems in parallel, each with only one reflecting coefficient vector. Specifically, for each $\bv_k$, the optimal solution can be obtained by maximizing the SINR $\gamma_k(\bv_k)$. To deal with the fraction-formed objective, 
\vspace{0.05cm}
\noindent we utilize FP in \cite{shen2018fractional} by introducing an auxiliary { {variable}} $\iota_k$. Then the the subproblem is equivalent to the following problem
\begin{small}
		\begin{align}
		&\hspace{-0.2cm}	 { \max_{\scriptstyle \bv_k,\iota_k} } \hspace{0.2cm}  2\Re \big\{ \iota_k^H   (h_{d,k}^{H} + \bb_{k}^{H} \bv_k )\big\} 	- 
		|\iota_k|^2 \big(    \noi_{n_{2}}  \bv_k^H \bQ_{1} \bv_k  + \noi_{z_{2,k}} \big) \label{pro:A3}\\
			& \hspace{0.cm} {\text{s.t.}} \hspace{0.4cm}   (\ref{pro:Ac}), \hspace{0.3cm} a_{k,n} \leq a_{\max},\hspace{0.1cm}  \forall n\in \mcal{N}, \tag{\ref{pro:A3}{a}} \label{pro:A3a} 
		\end{align}
\end{small}

{\noindent{\textls[0]{which can be solved by alternately optimizing $\iota_k$ and $\bv_k$ until the objective converges \cite{shen2018fractional}. Specifically, the optimal $\iota_k$ can be obtained by setting its first-order derivative to zero, which is given by}}}
\vspace{-0.1cm}
\begin{sequation}
\setlength\belowdisplayskip{0.cm}
	\begin{aligned}
		\iota_k = \frac{ (h_{d,k}^{H} + \bb_{k}^{H} \bv_k )}{ \noi_{n_{2}}  \bv_k^H \bQ_{1} \bv_k  + \noi_{z_{2,k}}} \vspace{-0.1cm} .
	\end{aligned} \label{eq:y_UE}
\end{sequation}\noindent
Then the optimal $\bv_k$ can be obtained by solving problem (\ref{pro:A3}) with the updated $\iota_k$, which is a quadratically constrained quadratic program (QCQP) problem and can also be solved with the interior-point algorithm. 

\vspace{-0.3cm}
\subsection{Complexity Analysis}
The computational complexity of the proposed algorithm for $(\pro_{UE})$ is given by {\small $\mathcal{O}\big( I_{AO} \ln (1/\epsilon) [ m \sqrt{4K+2N}  (4K+N^3+m (4K+N^2+N)+m^2 ) +K I_{FP}   (2N^{3.5}+KN^{2.5})] \big)$}  \cite{wang2014outage}, where $\epsilon$ is the solution accuracy and {\small $m=(N+1)(N+2)/2+2K$} is the number of variables in subproblem (\ref{pro:B1}). $I_{FP}$ and $I_{AO}$ are the number of iterations for the FP-based algorithm and the whole AO algorithm to reach the convergence, respectively.

\section{Proposed Solution for $(\pro_{UL})$ and $(\pro_{ST})$}
\vspace{-0.05cm}
\subsection{Proposed Solution for $(\pro_{UL})$}

Similarly, for the problem of employing UL-adaptive IRS beamforming, i.e., $(\pro_{UL})$, the original problem is first decomposed into two subproblems and then solved in an alternative manner. Note that the subproblem of optimizing $\tau_0,\tau_1,\pset$ and $ \bv_0$ with the given $\bv_1$ is exactly the same as that for (\ref{pro:B1}), which is thus omitted.

{\begin{spacing}{1.01}
Then we focus on optimizing $\bv_1$ with the given $\tau_0$, $\{\tau_k\},\pset$ and $ \bv_0$. Note that instead of the fraction-formed objective as for the second subproblem in  $(\pro_{UE})$, now we have to deal with the sum of logarithm function and fractions in the objective. Still, the FP method can be used to tackle this problem \cite{shen2018fractional}. Specifically, by introducing two auxiliary sets of variables $\chiset$ and $\ioset$, the subproblem is equivalent to \end{spacing}}
\vspace{-0.1cm}
\begin{small}
		\begin{align}
		&\hspace{-0.2cm}	{ \max_{\scriptstyle \bv_1 ,\chiset,\atop \scriptstyle \ioset } } \hspace{0.cm}
			\sum_{k=1}^K \omega_k \tau_1 \log_2 \big( 1+ \chi_k \big) -\omega_k \tau_1 \chi_k+ u_k(\bv_1 ,\chi_k ,\iota_k) \label{pro:B3}\\
			& \hspace{0.2cm} {\text{s.t.}} \hspace{0.4cm}  (\ref{pro:A2a}), \hspace{0.2cm} a_{1,n} \leq a_{\max},\hspace{0.2cm}  \hspace{0.2cm} \forall n\in \mcal{N}  , \tag{\ref{pro:B3}{a}} \label{pro:B3a} 
		\end{align}
\end{small}\noindent
where 
\begin{sequation}
	\begin{aligned}
		u_k&(\bv_1 ,\chi_k ,\iota_k) \teq  2\Re \big\{ \iota_k^H \sqrt{\omega_k \tau_1 ( 1+ \chi_k)p_k  } (h_{d,k}^{H} + \bb_{k}^{H} \bv_1 )\big\}  \\
		&-|\iota_k|^2 \big(  p_k |h_{d,k}^{H} + \bb_{k}^{H} \bv_1  |^2 + \noi_{n_{2}}  \bv_1^H \bQ_{1} \bv_1  + \noi_{z_{2,k}} \big).
	\end{aligned}\label{eq:uk}
\end{sequation}\noindent
{\textls[0]{{\begin{spacing}{1.01} \noindent 
Then the solutions of (\ref{pro:B3}) can be obtained by alternately optimizing $\chiset ,\ioset$ and $\bv_1$ until the objective converges \cite{shen2018fractional}. Specifically, the optimal $\bv_1$ can be obtained by solving the convex QCQP problem (\ref{pro:B3}) with the given $\chiset$ and $\ioset$. Whereas $\chiset$ and $\ioset$ are respectively updated by 
\end{spacing}}}}
\vspace{-0.0cm}
\begin{narsequation}
	\begin{aligned}
		\chi_k = \frac{ p_k | h_{d,k}^{H} + \bb_{k}^{H} \bv_1 |^2 }{ \noi_{n_{2}}  \bv_1^H \bQ_{1} \bv_1  + \noi_{z_{2,k}}}, \hspace{0.2cm} \forall k,
	\end{aligned}\label{eq:chi_UL}
\end{narsequation}\noindent
\begin{sequation}
	\begin{aligned}
		\iota_k = \frac{ \sqrt{\omega_k \tau_k ( 1+ \chi_k)p_k }(h_{d,k}^{H} + \bb_{k}^{H} \bv_1 )}{p_k | h_{d,k}^{H} \bb_{k}^{H} \bv_1  |^2 +   \noi_{n_{2}}  \bv_1^H \bQ_{1} \bv_1  + \noi_{z_{2,k}}}  , \hspace{0.2cm}\forall k.
	\end{aligned}\label{eq:iota_UL}
\end{sequation}\noindent

The computational complexity of the proposed algorithm for $(\pro_{UL})$ is {\small $\mathcal{O}\big( I_{AO} \ln (1/\epsilon) [ m \sqrt{4K+2N}  (4K+N^3+m (4K+N^2+N)+m^2 ) + I_{FP}   (2N^{3.5}+KN^{2.5})] \big)$ } \cite{wang2014outage}, where $m$ is the same as that for $(\pro_{UE})$.

\vspace{-0.2cm}
\subsection{Proposed Solution for $(\pro_{ST})$}
\vspace{-0.5cm}
{\textls[0]{{\begin{spacing}{1.01} 
For the problem of employing static IRS beamforming, i.e., $(\pro_{ST})$, first we optimize $\tau_0,\{\tau_k\}$ and $\pset$ with the given $\bv_0$, which can be formulated as the following convex problem 
\end{spacing}}}}
\vspace{-0.3cm}
\begin{small}
		\begin{align}
			&\hspace{-0.5cm} { \max_{\scriptstyle \tau_0,\{\tau_k\},\fset} } \hspace{0.5cm}
			\sum\nolimits_{k=1}^K \omega_k \tau_k \log_2 \big[ 1+ {\gamma_k(\bv_0) f_k }/{\tau_k} \big] \label{pro:3.1}\\
			& \hspace{0.cm} {\text{s.t.}} \hspace{1.2cm}  { f_k \leq \E_k ( \bv_0 ), \forall k, \hspace{0.2cm} (\ref{pro:B1c}), (\ref{pro:B1d})}. \tag{\ref{pro:3.1}{a}} \label{pro:3.1a}
		\end{align}
\end{small}\noindent

\vspace{-0.5cm}
{\textls[0]{{\begin{spacing}{1.01} 
For the subproblem of optimizing $\bv_0$ with the given $\tau_0,\{\tau_k\}$ and $\pset$, the objective can be reformulated similarly as given in (\ref{pro:B3}), except replacing $\bv_1$ with $\bv_0$, which is given by
\end{spacing}}}}
\vspace{-0.3cm}
\begin{small}
		\begin{align}
			& \hspace{-0.2cm}	{ \max_{\scriptstyle \bv_0 ,\chiset,\atop \scriptstyle \ioset } } \hspace{0.cm}
			\sum_{k=1}^K \omega_k \tau_1 \log_2 \big( 1+ \chi_k \big) -\omega_k \tau_1 \chi_k+ u_k(\bv_0 ,\chi_k ,\iota_k) \label{pro:3.2}\\
			& \hspace{0.2cm} {\text{s.t.}} \hspace{0.4cm}   (\ref{pro:Aa}), (\ref{pro:Ab}), (\ref{pro:A_STa}), (\ref{pro:A_STb}), \tag{\ref{pro:3.2}{a}} \label{pro:3.2a}
		\end{align}
\end{small}
\vspace{-0.3cm}

{\noindent
where $u_k(\bv_0,\chi_k ,\iota_k )$ is as given in (\ref{eq:uk}), $\chiset$ and $\ioset$ are updated by (\ref{eq:chi_UL}) and (\ref{eq:iota_UL}), respectively. }
Notice that the right-hand-side of the constraint (\ref{pro:Aa}) is a convex function with respect to $\bv_0$, thus is globally lower-bounded by its first-order Taylor expansion at the fixed point $\dv_0$, which is given by
\begin{sequation}
	\begin{aligned}
	&	P_\A | h_{d,k}^{H} + \bb_{k}^{H} \bv_0  |^2+ \noi_{n_1} \bv_0^H \bQ_{2,k} \bv_0 \geq  -\noi_{n_1} \dv_0^H \bQ_{2,k} \dv_0\\
		& + 2 \noi_{n_1} \Re \big\{ \dv_0^H \bQ_{2,k} \bv_0\big\}  -P_\A | h_{d,k}^{H} + \bb_{k}^{H} \dv_0  |^2 \\
		& + 2 P_\A \Re \big\{ (h_{d,k}^{H} + \bb_{k}^{H} \dv_0 ) (h_{d,k}^{H} + \bb_{k}^{H} \bv_0 )\big\} \teq q_k(\bv_0).
	\end{aligned}
	\label{eq:CR1_SCA}
\end{sequation}\noindent
By replacing {\small $	P_\A | h_{d,k}^{H} + \bb_{k}^{H} \bv_0  |^2+ \noi_{n_1} \bv_0^H \bQ_{2,k} \bv_0$} with $q_k(\bv_0)$ in constraint (\ref{pro:Aa}), the problem (\ref{pro:3.2}) turns into a QCQP problem and can be solved with the interior-point algorithm. 

To sum up, the computational complexity of solving $(\pro_{ST})$ 
{\begin{spacing}{1.2} 
\noindent with the proposed algorithm is {\small $\mathcal{O}\big( I_{AO} \ln (1/\epsilon) [(2K+1)^{3.5}  + I_{FP} 2\sqrt{2K+N}  (N^3+KN^2+N^2+KN)] \big)$}
 \cite{wang2014outage}.
\end{spacing}}

\section{Simulation Results}
We consider a WPCN with one HAP located at $(0, 0, 0)$ meter (m), one active IRS located at $(x_{\text{IRS}}, 0, 2)$ m, and $K$ devices which are randomly distributed in a circle centered at $(x_\text{{UE}}, 0, 0)$ m with a radius of 1 m.
We adopt the path loss model in \cite{wu2019intelligent} and the path loss exponent  is set to be 2.2 for $\bg$ and $\bh_{r,k}$, $\forall k$, while 3.5 for $h_{d,k}$, $\forall k$. For small scale fading, Rayleigh fading is adopted for $h_{d,k}$, $\forall k$. Whereas for $\bg$ and $\bh_{r,k}$, $\forall k$, Rician fading is adopted with a Rician factor of $10$. 
Other parameters are set as follows: $P_{\A}\nareq20$ dBm, $P_{\F}\nareq 5$ dBm, $\noi_{n1}\hspace{-0.15cm}=\hspace{-0.1cm}\noi_{n2}\nareq\noi_{z1}\nareq\noi_{z2}\nareq-90$ dBm, $T_{\max}\nareq1$ s, $x_{\text{IRS}}\nareq x_{\text{UE}}\nareq10$ m, $\eta  \hspace{-0.1cm}=\hspace{-0.05cm} 0.8$, $K\hspace{-0.1cm}=\hspace{-0.05cm} 4$ and $N\nareq 10$, if not specified otherwise.
{\textls[0]{{
For the purpose of comparison, we consider the following schemes.  1) \textbf{UE active:} the proposed algorithm for $(\pro_{UE})$; 2) \textbf{UL active:} the proposed algorithm for $(\pro_{UL})$; 3) \textbf{Static active:} the proposed algorithm for $(\pro_{ST})$; 
4) \textbf{UE passive:} employing a passive IRS with the user-adaptive beamforming via the algorithm proposed in \cite{wu2021irs}; 5) \textbf{UL/Static passive:} employing a passive IRS with the static beamforming via the algorithm proposed in \cite{wu2021irs}. }}}

\begin{figure*}[!t]
		\centering
		\subfigure[Sum throughput versus $P_\A$.]{
			\label{fig:PA_R}
			\includegraphics[width=0.488\columnwidth,height=0.44\columnwidth]{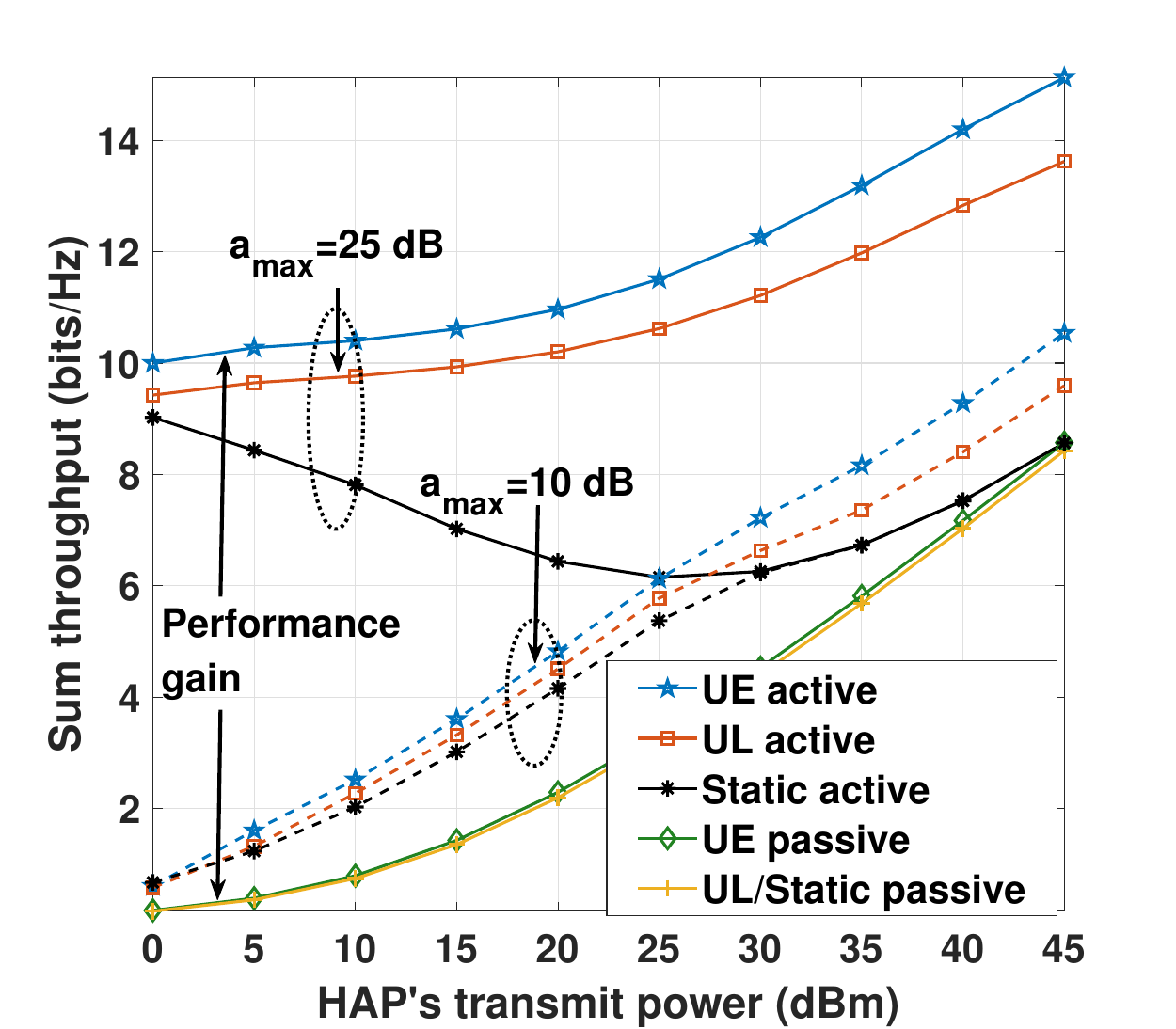}}
		\subfigure[Total energy consumption.]{
			\label{fig:PA_E}
			\includegraphics[width=0.488\columnwidth,height=0.44\columnwidth]{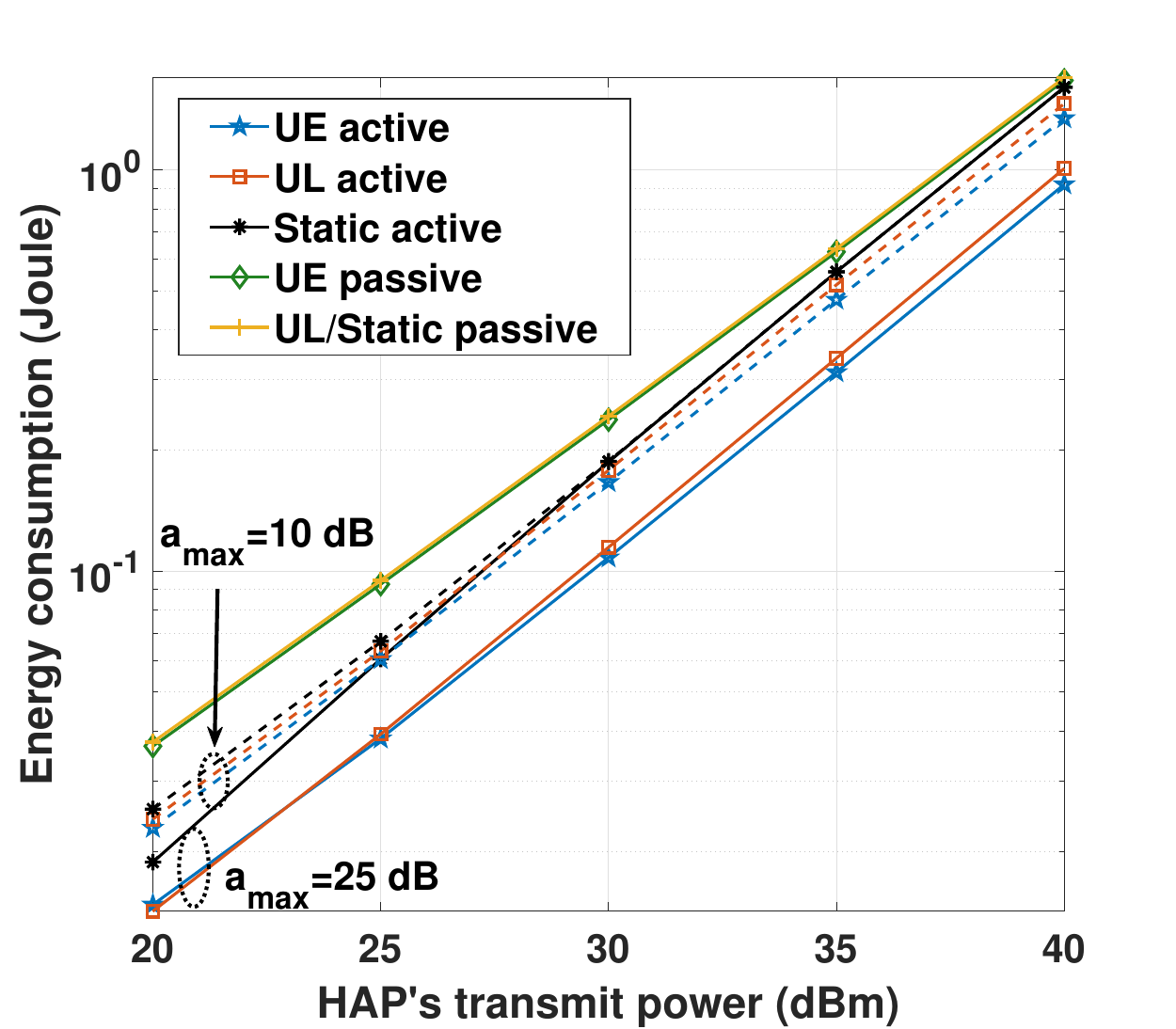}}
		\subfigure[IRS and UE cluster location.]{
			\label{fig:xIRS_UE_R}
			\includegraphics[width=0.488\columnwidth,height=0.44\columnwidth]{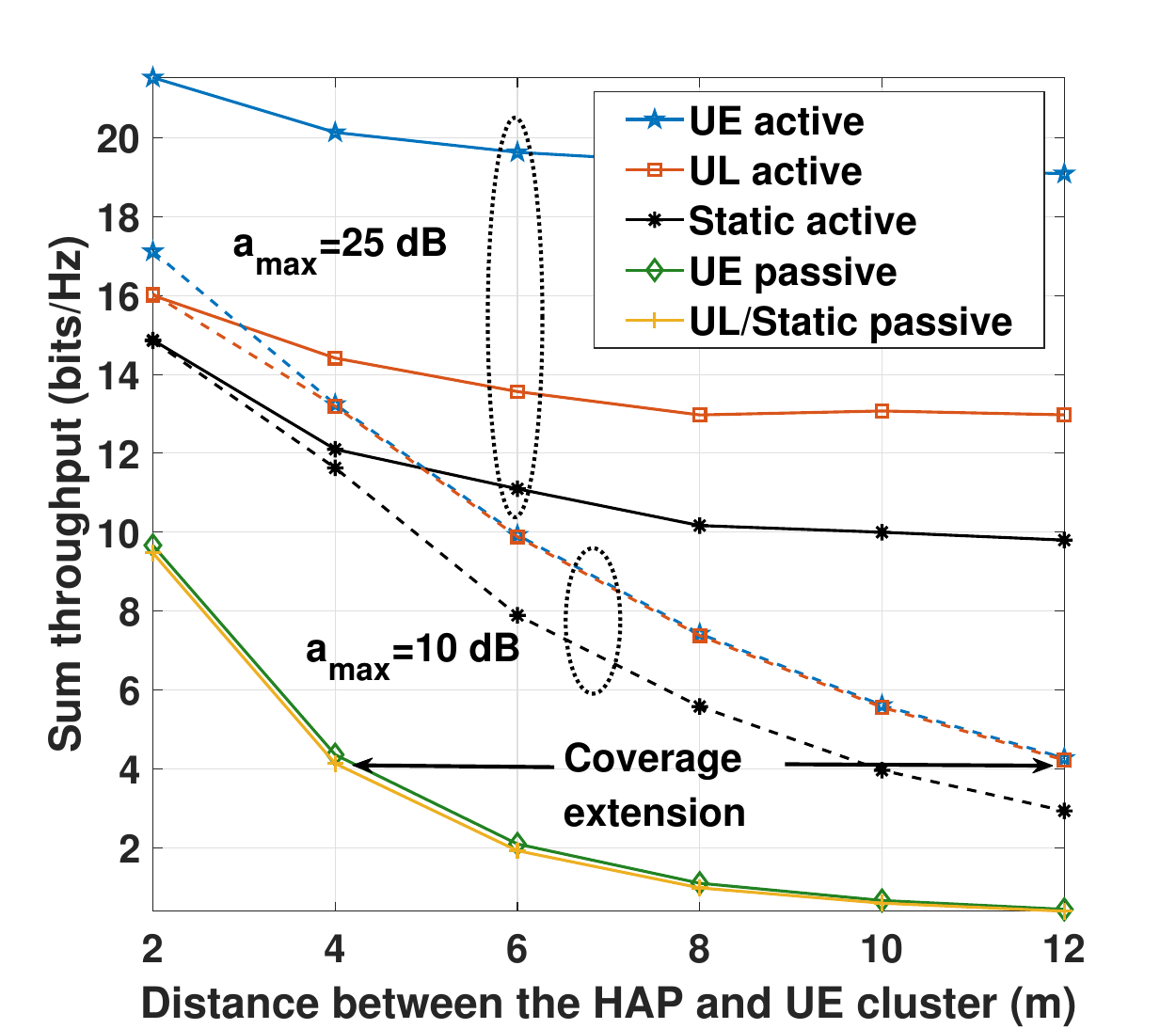}}
		\subfigure[IRS location.]{
			\label{fig:xIRS_R}
			\includegraphics[width=0.488\columnwidth,height=0.44\columnwidth]{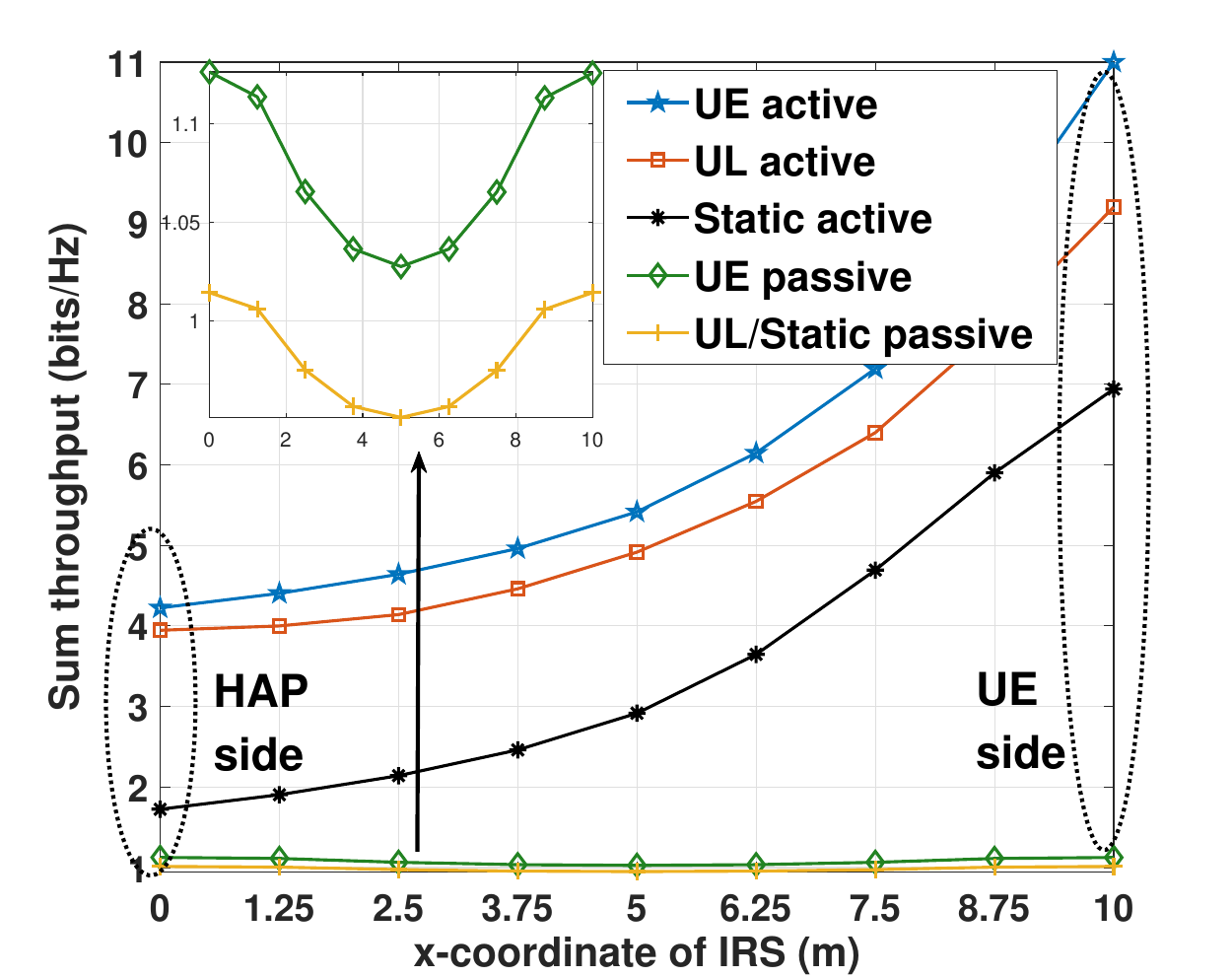}}
			\setlength{\abovecaptionskip} {-0.1cm}
			\caption{Performance under different setups.} 
	\end{figure*} 

\vspace{-0.21cm}
\subsection{Active or Passive IRS for WPCNs?}		

{\textls[0]{{
In Fig. \ref{fig:PA_R}, we plot the achievable sum throughput of the IRS-aided WPCN system versus the HAP's transmit power $P_\A$. For active IRS, we consider two cases with $a_{\max}$ = 10 and 25 dB \cite{amato2018tunneling}. It can be seen that employing the active IRS in WPCN performs much better than the traditional passive IRS-aided WPCN and this performance gain is more significant with larger constraint $a_{\max}$. 
}}}

As for the three different beamforming strategies, it is shown that for the active IRS, employing the UL-adaptive beamforming achieves an additional gain compared with employing static beamforming, which is different from the conclusion in \cite{wu2021irs}. This phenomenon results from the following two reasons. First, the active IRS amplifies the noise at the IRS, which can no longer be neglected in the optimization problem and brings an additional term in the constraint (\ref{pro:Aa}). {\textls[0]{{ Second, the IRS amplification power constraints for DL WET and UL WIT, i.e., (\ref{pro:Ab}) and (\ref{pro:Ac}), further break the symmetry of the DL and the UL channels. Besides, the performance gap is more significant when the constraint $a_{\max}$ is large, since $a_{\max}$ limits the amplification amplitude at the active IRS and thereby reduces the achievable performance gain.
This suggests that when $a_{\max}$ is large, we can employ more beamforming patterns at the active IRS to realize larger performance gain. Whereas it may not be necessary with the low $a_{\max}$.
}}}

Interestingly, we notice that employing static beamforming for the active IRS with $a_{\max}=25$ dB performs distinguishingly from the others, which presents an U-shaped curve as $P_\A$ increases. This is because the amplitude of the active IRS vector for DL WET, i.e., $|\bv_0|$ decreases as $P_\A$ increases due to the IRS amplification power constraints for DL WET, i.e., (\ref{pro:Ab}), which also influences the throughput for UL WIT since DL and UL share the same IRS patterns in this static setup.

 { {
Besides, we investigate the total energy consumption among the different schemes in Fig. \ref{fig:PA_E}, where the parameters are the same as those in Fig. \ref{fig:PA_R}. Note that the total energy consumption with the passive IRS is calculated as {\small $E_{passive}  \hspace{-0.1cm}=\hspace{-0.1cm} P_\A \tau_0$}. Whereas the total energy consumption with the active IRS is calculated as {\small $E_{active} \hspace{-0.1cm}=\hspace{-0.1cm} \tau_0 P_\A  \hspace{-0.0cm}+\hspace{-0.0cm} \tau_0 (P_\A \bv_0^H \bQ_{1} \bv_0 \hspace{-0.0cm}+\hspace{-0.0cm} \noi_{n_1} \bv_0^H \bv_0 ) \hspace{-0.0cm}+\hspace{-0.0cm} \sum_{k=1}^K \tau_k ( p_k \bv_k^H \bQ_{2,k} \bv_k \hspace{-0.0cm}+\hspace{-0.cm} \noi_{n_2} \bv_k^H \bv_k)$}. Surprisingly, employing passive IRS consumes more energy than that with the active IRS, due to a longer $\tau_0$ is needed for DL WET in passive IRS-aided WPCNs. {\textls[0]{{This validates the  superiority of utilizing active IRS compared with passive IRS, since the former one not only achieves higher throughput but also consumes less }}} transmit/amplifying energy under the given setups.  
}}

\vspace{-0.cm}
\subsection{Coverage Extension and Asymmetric Deployment}
\vspace{-0.cm}
In Fig. \ref{fig:xIRS_UE_R}, we plot the sum throughput versus the distance between the HAP and the center of the devices cluster, i.e., $x_{UE}$, where the IRS is deployed above the center of the devices cluster, i.e., {\textls[0]{{$x_{UE}\!=\!x_{IRS}$. It can be seen that, by using the active IRS, the transmission coverage improves greatly compared with employing passive IRS especially with a larger $a_{\max}$, which verifies the effectiveness of employing the active IRS in WPCNs.}}}

To gain more insights, we also evaluate the effect of the deployment of IRS. Specifically, we plot the sum throughput versus the x-coordinate of IRS, i.e., $x_{IRS}$, in Fig. \ref{fig:xIRS_R}, where $x_{UE}=10$ m. It can be seen that, different from the passive IRS-aided WPCNs, where the optimal IRS location is either above the HAP or the center of the devices cluster, the active IRS 
benefits an additional gain when deployed near the center of the devices cluster. This is because as the active IRS moves farther from the HAP, the received signal power at the IRS gets weaker, and thus according to (\ref{pro:Ab}), the active IRS can provide more amplification gain, which compensates for the attenuation caused by the double-fading effect. Therefore, more energy can be harvested at the wireless-powered devices to support the WIT. {\textls[0]{{This result indicates that for the WPCNs aided by an active IRS, it is better to deploy the IRS close to the devices.}}}

	\vspace{-0.cm}
	\section{Conclusions}
	\vspace{0.cm}
	In this paper, we investigated the joint beamforming and resource allocation optimization for an amplifying power limited active IRS-aided WPCN. 
	We considered three different beamforming setups and solved the corresponding WST maximization problems via SCA technique and FP method with the AO-based algorithm. 
	Numerical results not only demonstrated the significant superiority of adopting active IRS in WPCNs, but also validated the benefits of employing dynamic IRS beamforming. Particularly, it was found that by introducing active IRS, the WPCNs could achieve much higher throughput with less transmit/amplifying energy consumption compared to the passive IRS-aided WPCNs. Moreover, it was unveiled that the active IRS should be deployed near devices in practice.

	\vspace{-0.cm}
	
\bibliographystyle{IEEEtran}
\bibliography{IEEEabrv,main}
	
\end{document}